\documentclass[11pt]{article}
\usepackage{graphicx,amsmath,amsfonts,amssymb,fullpage,times,color}

\advance\parskip 2.5pt

\newtheorem{Remark}{Remark}[section]

\def\proof{\par{\it Proof}. \ignorespaces}
\def\endproof{{\ \vbox{\hrule\hbox{%
     \vrule height1.3ex\hskip0.8ex\vrule}\hrule }}\par}

\usepackage{rotating}
\newcommand{\rotxc}[1]{\begin{sideways}#1\end{sideways}}
\newcommand{\invert}[1]{\rotxc{\rotxc{#1}}}

\def\Y{\hbox{\invert{Y}}}

\numberwithin{equation}{section}

\numberwithin{figure}{section}

\let\trueint=\int
\let\truesum=\sum
\def\int{\mathop{\textstyle\trueint}\limits}
\def\sum{\mathop{\textstyle\truesum}\limits}

\let\<=\langle
\let\>=\rangle

\def\half{{\textstyle\frac12}}

\def\sf#1#2{{\textstyle\frac{#1}{#2}}}

\renewcommand\labelitemi{\ifmmode\circ\else$\circ$\fi}

\begin{document}

\title{Construction of KP solitons from wave patterns}
\author{Sarbarish Chakravarty$^1$ and Yuji Kodama$^2$ \\[1ex]
\small\it\
$^1$Department of Mathematics, University of Colorado, Colorado Springs, CO 80933 \\
\small\it\
$^2$ Department of Mathematics, Ohio State University, Columbus, OH 43210}
\date{}
\maketitle
\begin{abstract}
We often  observe that waves on the surface of shallow water form complex web-like patterns.
They are examples of nonlinear waves, and these patterns are generated 
by nonlinear interactions among several obliquely propagating waves.
  In this note, 
we discuss how to construct an exact soliton
solution of the KP equation from such web-pattern of shallow water wave.
This can be regarded as an ``inverse problem" in the sense that by
measuring certain metric data of the solitary waves in the given
pattern, it is possible to construct an exact KP soliton solution which can
describe the non-stationary dynamics of the pattern.
\end{abstract}

\thispagestyle{empty}

\section{Introduction}

\begin{figure}[h!] 
\begin{center}
\includegraphics[width=5cm]{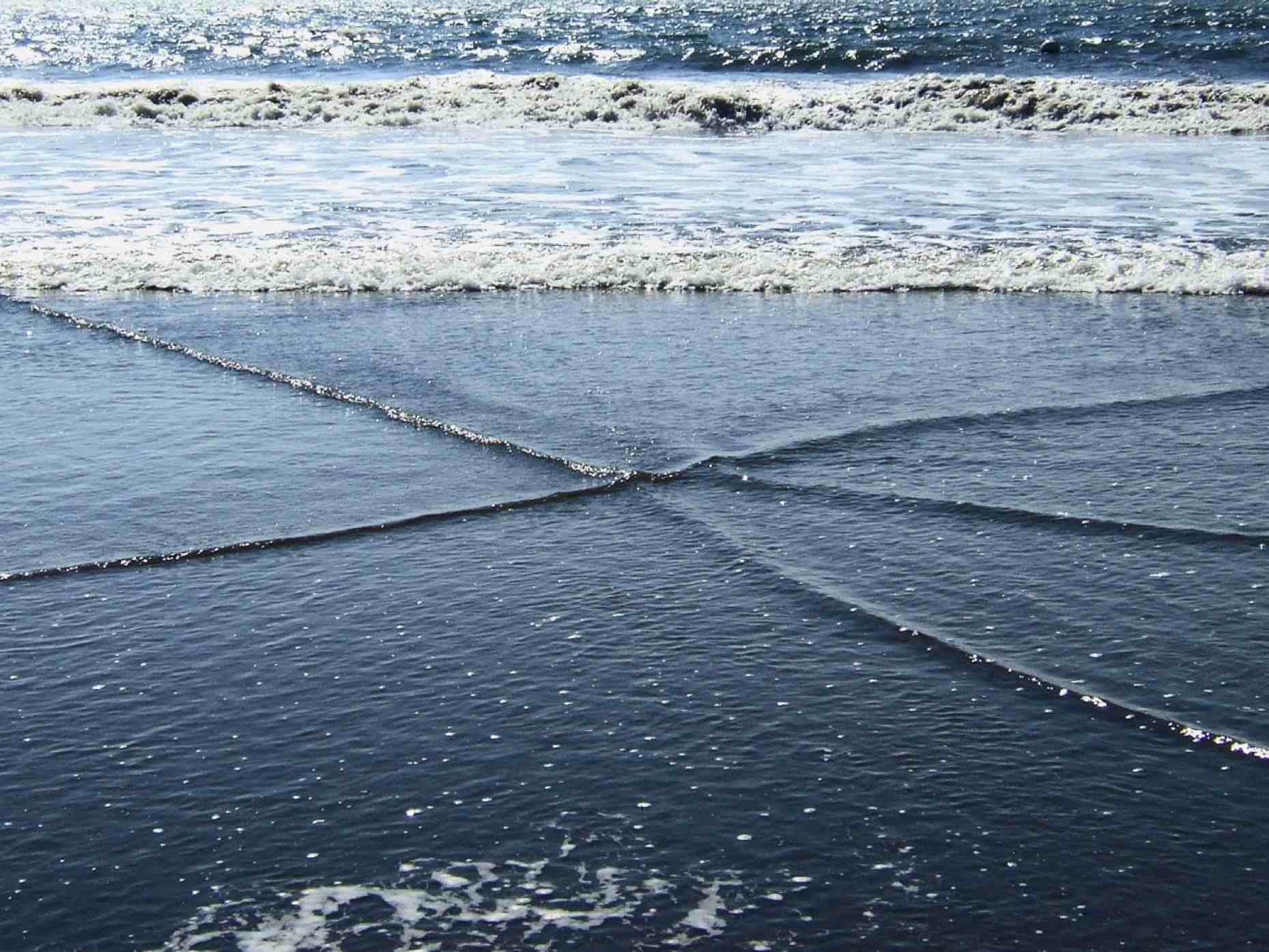} \hskip2cm
\includegraphics[width=5cm]{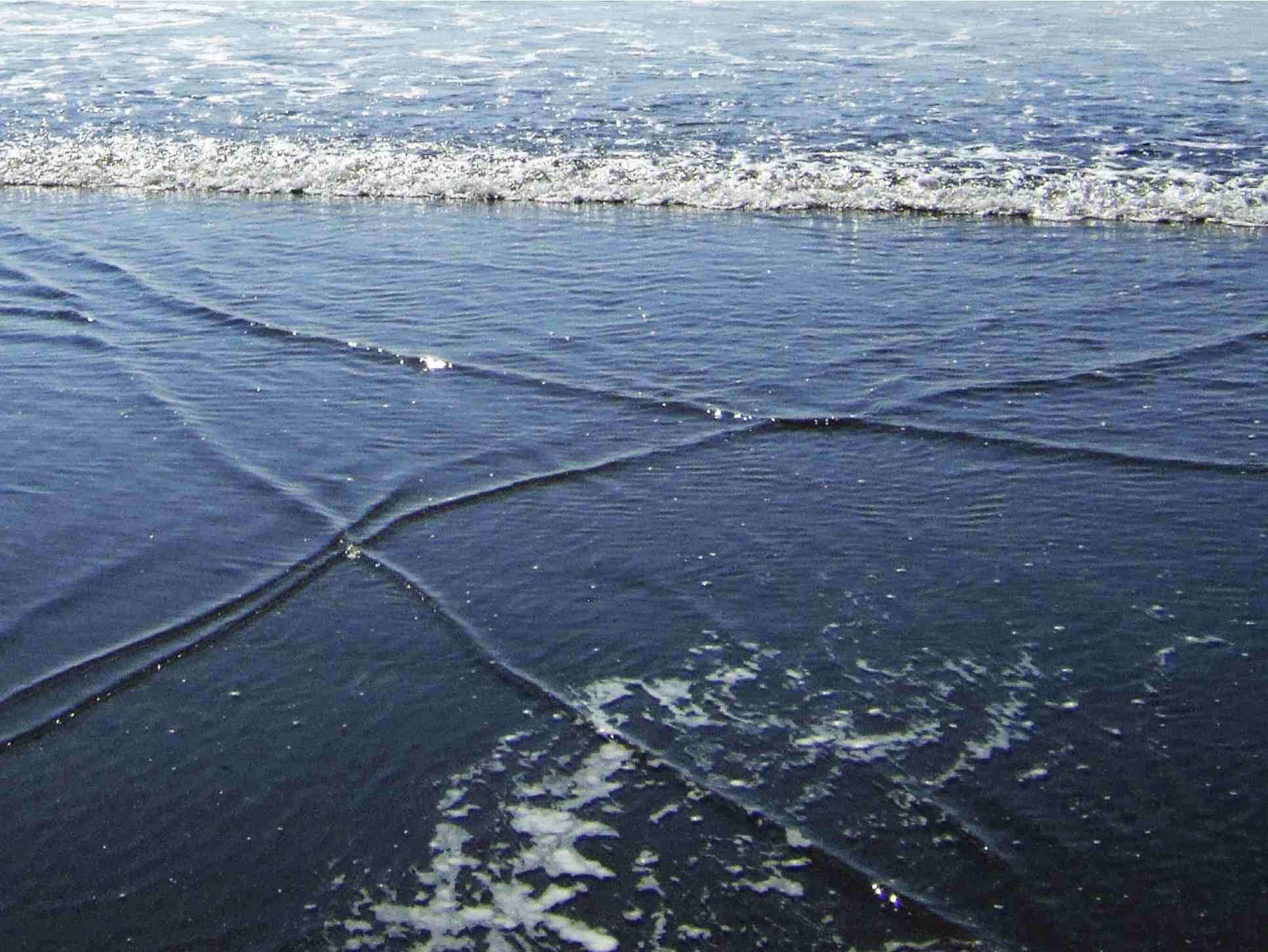}
\end{center}
\caption{Wave patterns on a beach at Nuevo Vallarta, Mexico. Photographs by M. J. Ablowitz}
\label{fig:f120015}
\end{figure}

At any beach with flat or nearly flat bottom, one often observes  
interesting wave patterns as shown in Figure \ref{fig:f120015}.
In this paper, we provide a mathematical analysis for those patterns
(non-stationary in general) based on the
Kadomtsev-Petviashvili (KP) equation \cite{KP:70} which is given by
\begin{equation}\label{kp}
(4u_t+6uu_x+u_{xxx})_x+3u_{yy}=0.
\end{equation}
Here $u=u(x,y,t)$ represents the normalized wave amplitude at the point $(x,y)$ in
the $xy$-plane for fixed time $t$.

From a physical perspective, the KP equation is derived from the three-dimensional 
Euler equations for an irrotational and incompressible fluid under the
assumptions that it describes the propagation of small amplitude, long wavelength,
uni-directional waves with small transverse variation 
(i.e., quasi-two-dimensional waves). An example of such wave phenomena 
observed in nature is the surface wave patterns in shallow water on long, 
flat beaches as shown in Figure~\ref{fig:f120015}. 
On a smaller scale, such wave patterns 
can be easily demonstrated in table-top experiments~\cite{KOT:09} 
(see also movies available at the website http://www.needs-conferences.net/2009/), 
as well as recreated in more
accurate water tank experiments in the laboratory~\cite{LYK:11,Y:13}.

The KP equation admits an important class of solitary
wave solutions that are regular, non-decaying and localized along distinct
lines in the $xy$-plane. These are known as the line-soliton solutions which
have been studied extensively in recent years by the authors who have provided
a complete classification of these solutions using geometric and combinatorial
techniques~\cite{CK:08, CK:08b, CK:09, K:10}.  In principle, these solutions may
have arbitrary number of asymptotic line solitons in the far-field and a complex
interaction pattern of intermediate solitons resembling a web-like structure
in the near-field region. Because of this, we sometimes call the line-soliton
solutions the \emph{web-solitons}.  Several simple yet exact web-solitons, 
for example, the Y-shape soliton and several other solutions with two asymptotic 
solitons for $|y| \gg 0$, have been 
experimentally demonstrated. Indeed, in recent laboratory wave tank experiments 
by Harry Yeh's group at Oregon State University surface wave patterns generated 
by solitary wave interactions which are very good approximations of the exact 
KP solutions, have been observed~\cite{LYK:11,Y:13}.

The purpose of this note is to study the {\it inverse problem}, i.e., to
construct an exact line-soliton solution of the KP equation that approximates
an observed wave pattern. Specifically, we consider the wave patterns observed
in shallow water such as in Figure \ref{fig:f120015}. We assume that these 
patterns satisfy the assumptions stated above, necessary to derive
the KP equation from the Euler equations. Then the inverse problem determines the
information required to construct the exact line-soliton solution from the wave 
pattern data consisting of the amplitude and slopes of the solitary waves,
and the locations of those waves on the $xy$-plane for fixed $t$. 
We emphasize at the very outset that in this note we limit our considerations 
only to the comparison between the wave patterns and the exact KP theory, that is,
we do not take into account any higher order corrections to the KP equation
due to large amplitude, finite angle (departure from quasi-two-dimensionality),
uneven bottom or any such physical perturbations. In this sense, the inverse
problem considered here should be regarded as a leading order approximation.

Recent interest in the study of line-soliton
solutions of the KP equation has generated a data bank of photographs and video 
recordings of patterns formed by interacting small amplitude solitary waves in shallow 
water on flat beaches, see for example, a recent article
by Ablowitz and Baldwin~\cite{AB:12} and additional resources at
the authors' websites~\cite{AB:12a}.  
In this paper we propose an algorithm to construct the line-soliton solutions
approximating those given wave patterns.
Modeling of shallow water wave patterns by the line-soliton solutions are 
carried out in a ``qualitative'' fashion in the current literature 
(see e.g. \cite{AB:12, PG:00, P:02, So:06}), and at times, using ad hoc methods. 
Most such studies consider the KP $N$-soliton solutions given by the well-known 
Hirota  formula~\cite{H:04}. But even for $N=2$, this formula is not adequate 
to describe the rich variety of similar interaction patterns revealed
by the KP theory~\cite{CK:08, CK:09} since the 2-soliton solution obtained
from the Hirota formula only gives a \emph{stationary} X-shape pattern. In contrast, 
our algorithm makes concrete \emph{quantitative} use of the measurements from a 
given wave pattern to obtain the explicit analytical form of the KP line-soliton 
solution which can describe the non-stationary dynamics of the pattern.
 
It is important to recognize that the resonant interaction among solitary
waves plays a fundamental role in the formation of surface wave patterns
that can be approximated by the KP solitons.
It was Miles~\cite{M:77a, M:77b} who first identified the resonant interaction in 
KP solitons when two asymptotic line solitons of an X-shape 2-soliton
solution, referred here as the X-soliton, interact
obliquely at a certain critical angle, and a third soliton is created 
to make a Y-shape pattern. It turns out that such Y-shape wave-form 
is an exact solution of the KP equation, and is referred to
as Y-soliton (see also \cite{NR:77}). Subsequently, more general type of 
resonant and partially-resonant line-soliton solutions have been
discovered (see e.g.,~\cite{BK:03,K:04,BC:06}). In this paper, we demonstrate 
using explicit examples of shallow water wave patterns that the resonant 
line-soliton solutions of the KP equation can be used to approximate such patterns 
and their dynamics fairly well.

The paper is organized as follows:
In Section 2, we provide some basic background for the one-soliton solution
of the KP equation, as well as the resonant Y-soliton and the non-resonant X-soliton
solutions. Then in Section 3
we describe an algorithmic method to construct an exact KP soliton solution
from a given wave pattern.
In Section 4, we illustrate our algorithm via several explicit examples of actual shallow water
wave patterns,
and demonstrate that the dynamics given by the exact solutions 
are in good agreement with the observations.
We finally give some concluding remarks in Section 5.


\section{The KP solitons}
Here we briefly give the background information on the KP solitons necessary to
this paper and a remark on some particular solutions (see \cite{CK:08,CK:09,K:10} for the details).

It is customary to prescribe the solutions of the KP equation \eqref{kp} as
\begin{equation}
u(x,y,t) = 2(\ln \tau)_{xx} \,,
\label{u}
\end{equation}
in terms of the $\tau$-function $\tau(x,y,t)$ (see e.g. \cite{H:04}).  For the soliton solutions,
the $\tau$-function is defined via
\begin{itemize}
\item[(i)] $M$ distinct real 
parameters $\{k_1,k_2,\ldots,k_M\}$ with the ordering $k_1 < k_2 < \cdots <k_M$, and 
\item[(ii)] an $N \times M$ real matrix $A$ of full rank with $N < M$.  
\end{itemize}
The explicit form of the $\tau$-function is as follows:
\begin{equation}
\tau(x,y,t)=\sum_{I}\Delta_I(A)\,E_I(x,y,t)\,,
\label{tau}
\end{equation}
where the sum is over all (ordered) $N$-element subsets of $[M]:=\{1,\ldots, M\}$,
denoted by $\binom{[M]}{N}$ and $|I|=N$ denotes the number of elements of $I$, i.e., $I=\{i_1<\cdots<i_N\} \in \binom{[M]}{N}$.
The coefficient $\Delta_I(A)$ is the $N \times N$ minor of the matrix $A$ with 
the column set $I $, and $E_I(x,y,t) := K_I\exp\Theta_I(x,y,t)$ where
\begin{equation}\label{KT}
K_I=\prod_{l>m}(k_{i_l}-k_{i_m}), \qquad \quad 
\Theta_I(x,y,t)=\sum_{m=1}^N(k_{i_m}x+k_{i_m}^2y-k_{i_m}^3t).
\end{equation}
 The soliton solution $u(x,y,t)$ is
regular if and only if  $\Delta_I(A) \ge 0$
for all $I \in\binom{[M]}{N}$~\cite{KW:11, KW:12}. In this case, the 
matrix $A$ is called a {\it totally non-negative} matrix.  

In our previous works~\cite{CK:08,CK:09},
it was shown that the general soliton solution given
by equations \eqref{u} and \eqref{tau} consists of $N$ line-solitons 
as $y \gg 0$ and $M-N$ line-solitons as $y \ll 0$. 
Each of those asymptotic solitons is uniquely parametrized by a pair
of distinct $k$-parameters $\{k_i, k_j\}$ for $i < j $. 
We let $[i,j]$ denote the index pair for this soliton.
Furthermore, the index pair $[i,j]$ is uniquely characterized by a 
map $\pi$ such that $\pi(i)=j$ if $[i,j]$ labels an asymptotic soliton
for $y \gg 0$, and $\pi(j)=i$ if $[i,j]$ labels an asymptotic soliton
for $y \ll 0$. The map $\pi$ turns out to be fixed-point
free permutation of the index set $[M]$ known as {\it derangement}
which is conveniently represented by a chord diagram as shown in examples below.
Thus the soliton solution generated by the $\tau$-function \eqref{tau} is represented 
by the chord diagram associated to the derangement $\pi$.

We briefly describe below some details of the one-solitons, the Y-solitons
and the X-solitons which form the building blocks of a web-soliton
solution of the KP equation.  Those One-, Y- and X-solitons are stationary solutions
of the KP equation.  We emphasize that 
the web-solitons consisting of those solutions are no longer stationary.

\subsection{One soliton solution}
\begin{figure}[h!]
\begin{center}
\includegraphics[width=3.8cm]{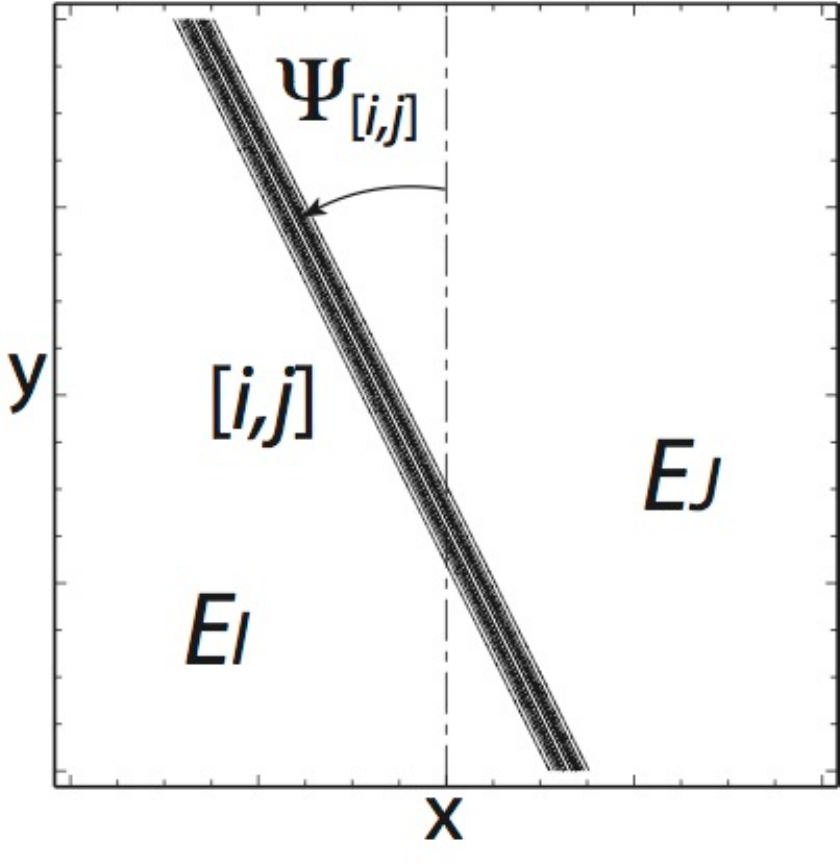} \hskip2cm 
\raisebox{0.5cm}{\includegraphics[width=3.6cm]{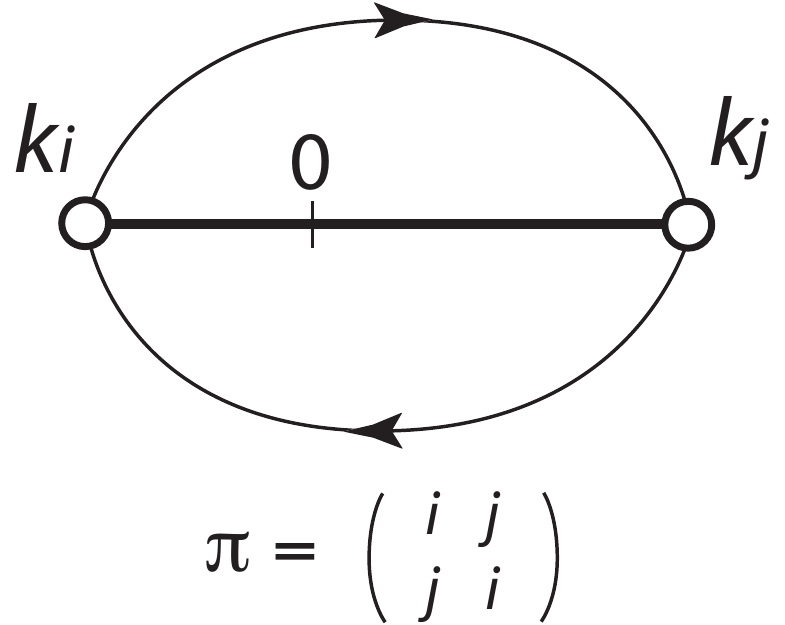}}
\end{center}
\caption{Contour plot  and the chord diagram representation of the $[i,j]$-soliton}
\label{fig:1soliton}
\end{figure}

Asymptotic analysis of the $\tau$-function in \eqref{tau} reveals that the 
solution $u(x,y,t)$  is exponentially vanishing in regions 
of the $xy$-plane
where a single exponential $E_I$ is dominant over all other exponentials 
in \eqref{tau} while it is localized along certain lines where a pair of
dominant exponentials $E_I, E_J$ are in balance. Each such line is labeled
by an index pair $[i, j]$ as explained below. The solution 
$u(x,y,t)$ along this line is approximated by a one-soliton solution,
and will be referred to as the $[i,j]$-soliton throughout this note. 
Along this line $[i,j]$, the $\tau$-function can be locally approximated by
\[
\tau~ \approx ~\Delta_I(A)E_I+\Delta_J(A)E_J\,,
\]
where $I=I_0\cup \{i\}$ and $J=I_0\cup\{j\}$ 
with $|I_0|=N-1$ (see \cite{CK:09}). Then near the line $[i,j]$,
the solution $u=2(\ln\tau)_{xx}$ has the form of a one-soliton, 
\begin{equation}\label{soliton}
u\,\approx \,A_{[i,j]}\,\mathrm{sech}^2\,\Theta_{[i,j]}\,, 
\end{equation}
localized along the line $[i,j]$ given by $\Theta_{[i,j]}=0$ where
\begin{align}
\Theta_{[i,j]}:=\frac{1}{2}(\Theta_I-\Theta_J)&=
\frac{1}{2}(k_i-k_j)\left(x+\tan\Psi_{[i,j]}y-C_{[i,j]}t+x_{[i,j]}\right) 
\label{theta} \\
x_{[i,j]}&=\frac{1}{k_i-k_j}\ln\frac{\Delta_I(A)K_I}{\Delta_J(A)K_J} \nonumber \,.
\end{align}
(recall \eqref{KT} for the formulae $K_I$ and $\Theta_I$).
The soliton amplitude $A_{[i,j]}$, slope $\Psi_{[i,j]}$ and velocity
in the positive $x$-direction $C_{[i,j]}$, are defined respectively by 
\begin{align}
&A_{[i,j]}=\frac{1}{2}(k_i-k_j)^2, \qquad\tan\Psi_{[i,j]}={k_i+k_j},
\label{ij}\\
&C_{[i,j]}=k_i^2+k_j^2+k_ik_j>0 \nonumber .
\end{align}
A contour plot of this one-soliton solution is shown in Figure~\ref{fig:1soliton}.
Note that the angle $\Psi_{[i,j]}$ is measured counterclockwise from the $y$-axis,
and $-\sf{\pi}{2}<\Psi_{[i,j]}<\sf{\pi}{2}$. Since the parameter $k_i$ appears only in 
the phase $\Theta_I$ of the
exponential $E_I$ while $k_j$ appears only in $\Theta_J$ of the exponential
$E_J$, the line $[i,j]$ can be viewed as representing a permutation of the index set 
$\{i, j\}$. That is, the parameters $k_i$ and $k_j$ are exchanged during the
transition from one dominant exponential to the other by crossing the $[i,j]$-soliton.
The chord diagram in Figure~\ref{fig:1soliton} depicting the exchange 
$k_i \leftrightarrow k_j$ represents the permutation 
$\pi=\left(\begin{smallmatrix}i&j\\j&i\end{smallmatrix}\right)$ 
of the index set $\{i, j\}$.

\subsection{Y-soliton solution}
\begin{figure}[h!]
\begin{center}
\raisebox{1.4in}{(a)}~\includegraphics[width=4cm]{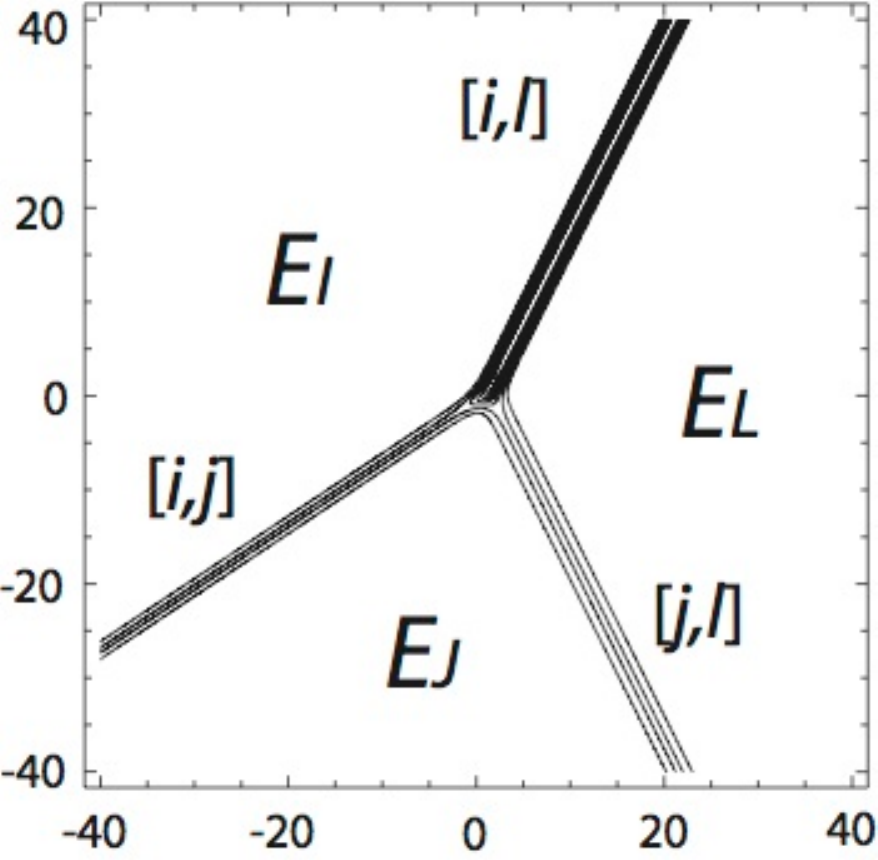}
\hskip0.8in\raisebox{1.4in}{(b)}~\includegraphics[width=4cm]{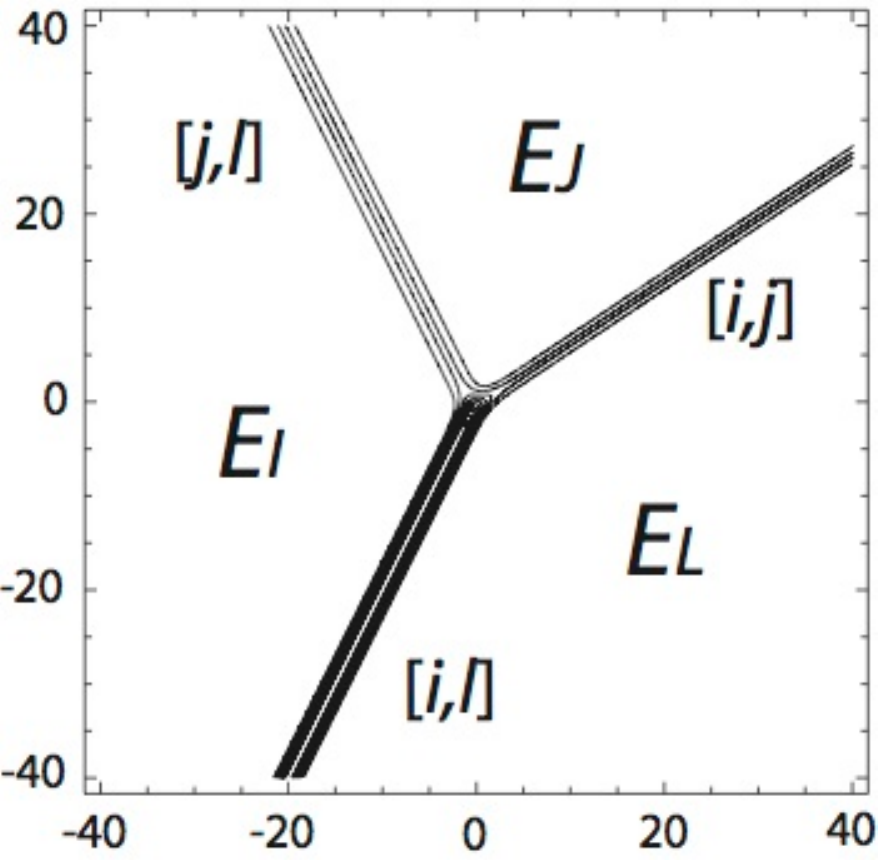}
\end{center}
\caption{Contour plots of the Y-solitons.  Resonant soliton triplet $[i,j],[j,l]$ and $[i,l]$.}
\label{fig:Ysoliton}
\end{figure}

\begin{figure}[h!]
\begin{center}
\raisebox{0.6in}{(a)}~\includegraphics[width=4cm]{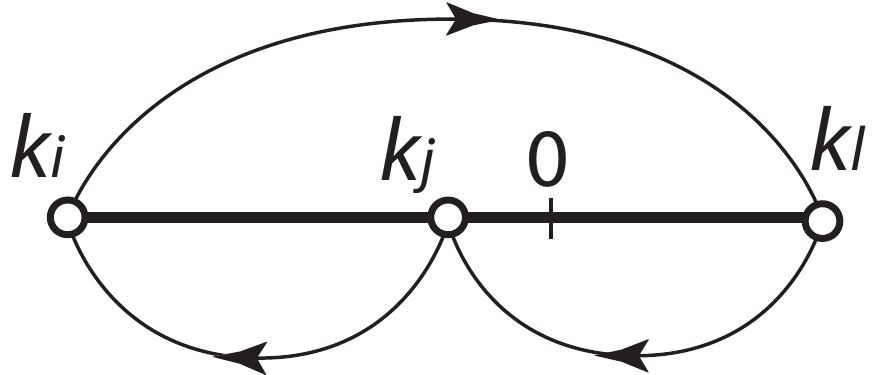}
\hskip0.8in\raisebox{0.6in}{(b)}~\includegraphics[width=4cm]{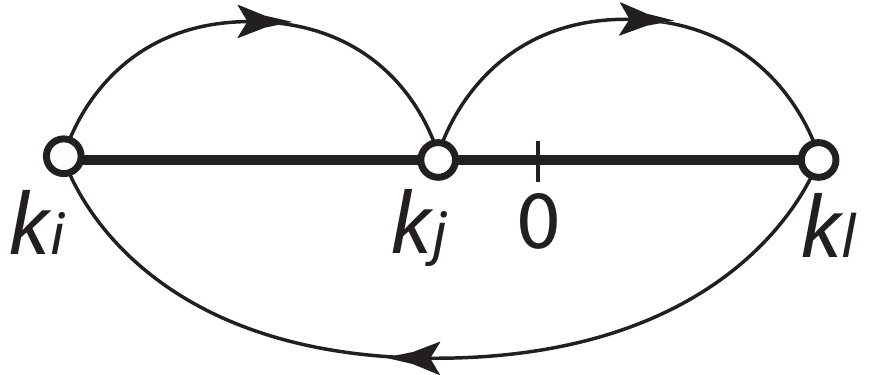}
\end{center}
\caption{Chord diagrams for the Y-solitons  in Figure \ref{fig:Ysoliton}.
The horizontal line gives the coordinate of the $k$-parameters,
i.e. $k_i<k_j<0<k_l$ in those diagrams.}
\label{fig:Ychord}
\end{figure}

Here we briefly describe the Y-soliton which is a system of three 
line solitons labeled $[i,j]$, $[j,l]$ and  $[i,l]$ with $i<j<l$,
interacting at a trivalent vertex.  Writing each of the solitons in the form
of a traveling wave $u=\Phi(\mathbf{K}_{[a,b]}\cdot\mathbf{r}-\Omega_{[a,b]}t)$ with 
$\mathbf{r}=(x,y)$, the wave-vector $\mathbf{K}_{[a,b]}=\half(k_b-k_a,k_b^2-k_a^2)$ 
and the frequency $\Omega_{[a,b]}=\half(k_b^3-k_a^3)$, 
the soliton triplet satisfy the resonant conditions
\[
\mathbf{K}_{[i,j]}+\mathbf{K}_{[j,l]} = \mathbf{K}_{[i,l]}\,, \quad \qquad 
\Omega_{[i,j]}+\Omega_{[j,l]} = \Omega_{[i,l]} \,.
\]
Near the trivalent vertex, the $\tau$-function has the form 
\[ 
\tau\approx ~\Delta_I(A)E_I+\Delta_J(A)E_J +\Delta_L(A)E_L \,.
\]
There are two cases for the index sets $\{I,J,L\}$ leading to the
contour plots in Figures~\ref{fig:Ysoliton}. 
\begin{itemize}
\item[(a)] The index sets are given by
$I=I_0\cup\{i\}, J=I_0\cup\{j\}$ and $L=I_0\cup\{l\}$.
In this case the $[i,l]$-soliton is above the trivalent vertex while
the solitons $[i,j]$ and $[j,l]$ appear below it to form a \Y-shape as shown 
in Figure~\ref{fig:Ysoliton}(a). Following the transitions of the dominant 
exponentials clockwise in Figure~\ref{fig:Ysoliton}(a), one recovers the 
permutation $\pi= \left(\begin{smallmatrix}i&j&l\\ l&j&i\end{smallmatrix}\right)$
given by the three line solitons $[i,j], \, [j,l], \,[i,l]$. The chord diagram
corresponding to this permutation is shown in Figure~\ref{fig:Ychord}(a).  
\item[(b)] The index sets are  
$I=J_0\cup\{i,j\}, J=J_0\cup\{i,l\}$ and $L=J_0\cup\{j,l\}$. The solution
in this case is related to that of Case (a) by an inversion $(x,y,t) \to (-x,-y,-t)$
so that the solitons $[i,j]$ and $[j,l]$ appear above the trivalent vertex 
and the $[i,l]$-soliton appears below it. The line solitons for this Y shape
represent the permutation $\pi= \left(\begin{smallmatrix}i&j&l\\j&l&i\end{smallmatrix}\right)$
as shown by the chord diagram in Figure~\ref{fig:Ychord}(b).
\end{itemize}
 
Due to the ordering $k_i<k_j<k_l$,
the $[i,l]$-soliton has the largest amplitude
$A_{[i,l]}=\half(k_l^2-k_j^2)$, and the $[j,l]$-soliton has the largest slope 
$\tan\Psi_{[j,l]}=k_j+k_l$,
while the $[i,j]$-soliton has the smallest slope. Note that all these
information can be gathered directly from the chord diagrams presented
in Figure \ref{fig:Ychord}, which is one of the main reasons we use the
chord diagrams in our analysis. 
One should also note that the $k$-parameters in the 
Y-soliton are uniquely determined only from the slopes of the three solitons. 
Recall that the slopes of the  solitons $[i,j], \, [j,l], \,[i,l]$ are respectively given by
\[
\tan\Psi_{[i,j]} = k_i+k_j, \quad \tan\Psi_{[j,l]} = k_j+k_l, \quad
\tan\Psi_{[i,l]} = k_i+k_l \,.
\]
 Then one can compute the values of the $k$-parameters as
\begin{align}
k_i&=\half \left(\tan\Psi_{[i,l]}+\tan\Psi_{[i,j]}-\tan\Psi_{[j,l]}\right), \nonumber\\
k_j&=\tan\Psi_{[i,j]}-k_i, \label{ijl} \\
k_l&=\tan\Psi_{[i,l]}-k_i \nonumber \,.
\end{align}

\subsection{X-soliton solution and the phase shift}\label{sec:X}
\begin{figure}[h!]
\begin{center}
\includegraphics[width=5.5cm]{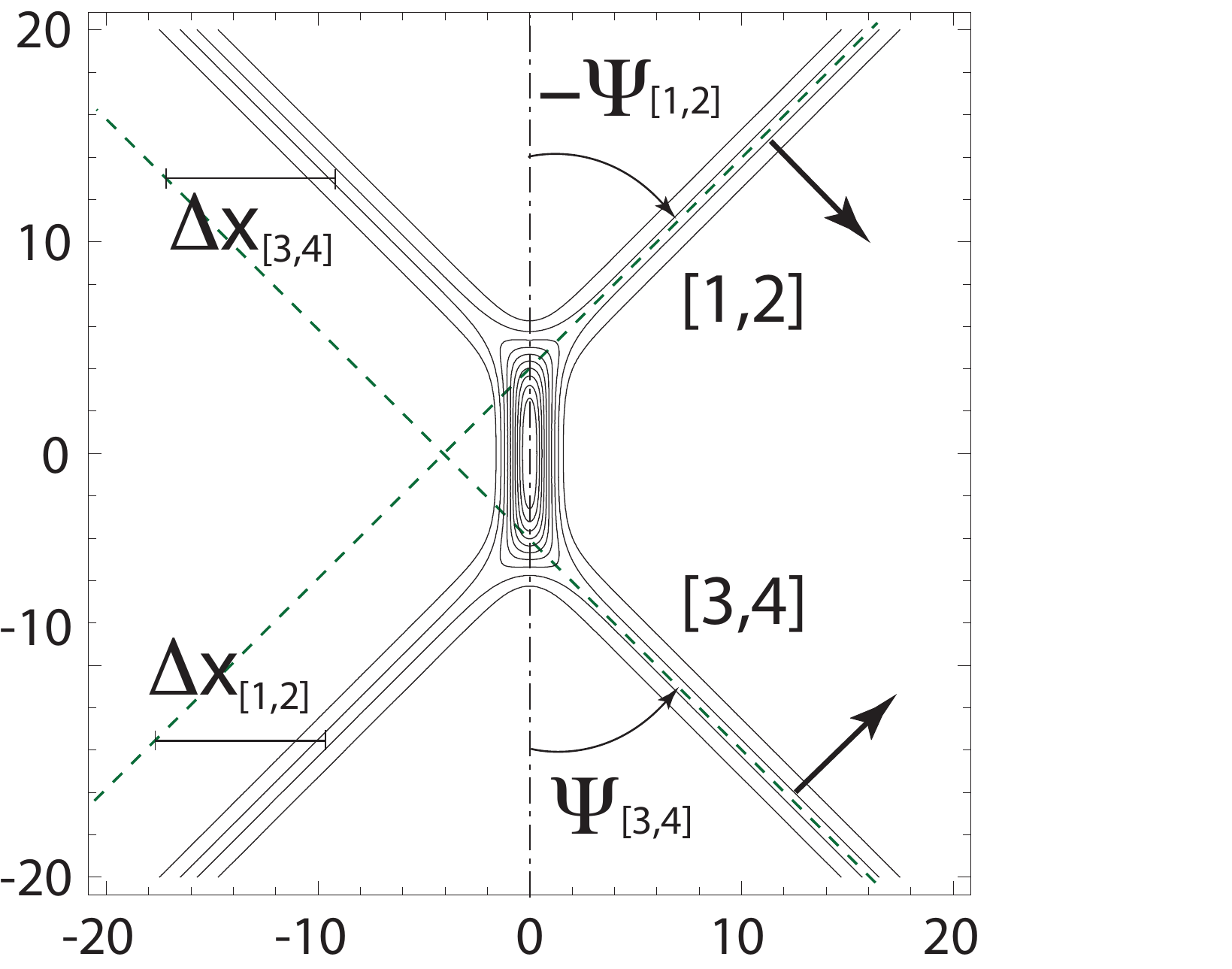} \quad
\raisebox{0.3 in}{\includegraphics[width=5.5cm]{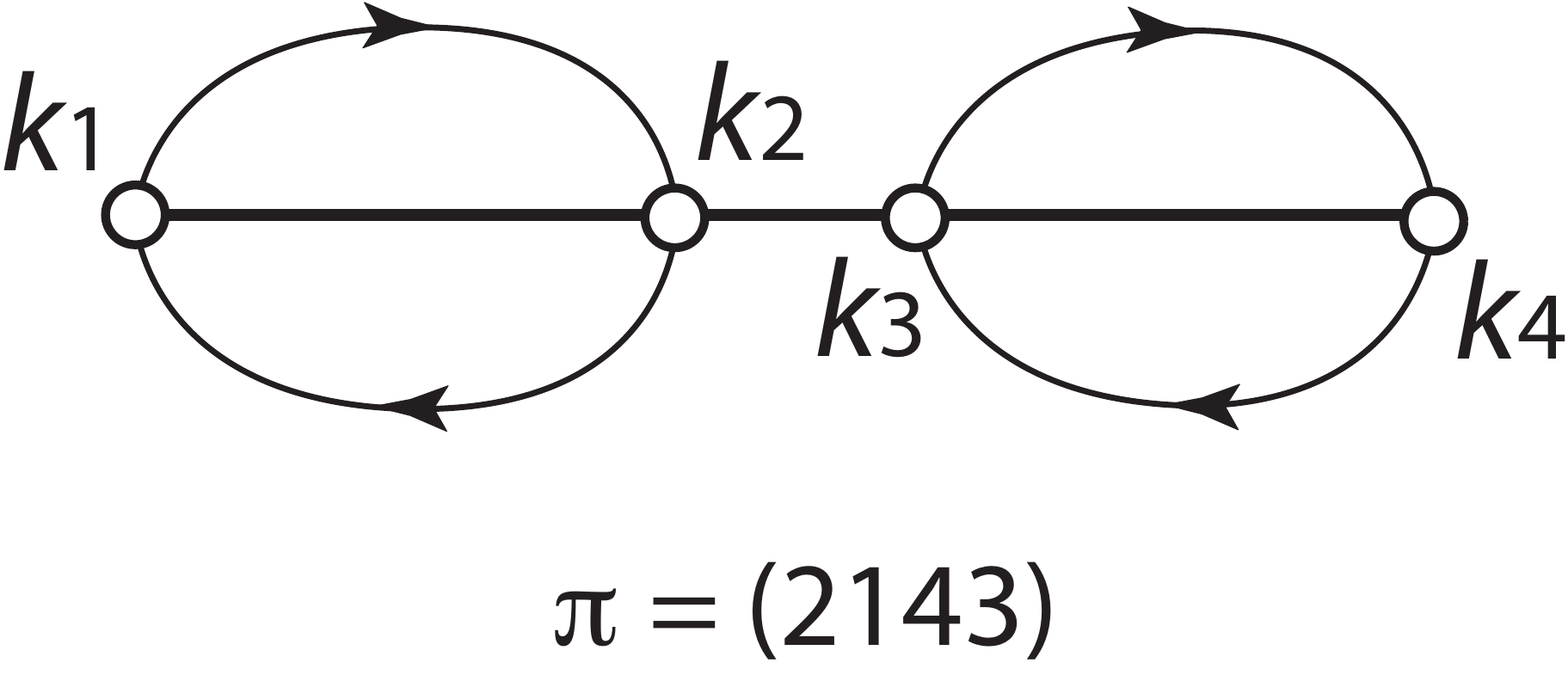}}
\end{center}
\caption{X-type soliton interaction with phase shifts $\Delta x_{[i,j]}$ and the chord diagram.}
\label{fig:PS}
\end{figure}

Two line-solitons can form an X-vertex as a result of interaction.
In this case, the interaction is \emph{non-resonant}, i.e. those two solitons
do not generate a third soliton with the resonant condition, and
the wave pattern is \emph{stationary}.
The pair of line solitons can either be labeled as  $[i,j]$- and $[k,l]$-solitons
or, $[i,l]$- and $[j,k]$-solitons with $i<j<k<l$.
In terms of the chord diagram, those are described by two non-crossing 
closed chords (loops) connecting the index pairs $\{ [i,j],[k,l]\}$ or $\{ [i,l], [j,k]\}$
(see the example in Figure~\ref{fig:PS}).
Near the X-vertex, the $\tau$-function has the following form containing four exponential terms,
\[
\tau~\approx~ \Delta_I(A)E_I+\Delta_J(A)E_J+\Delta_L(A)E_L+\Delta_M(A)E_M,
\]
where $I,J,L$ and $M$ have $N-2$ common indices (see \cite{CK:09} for the details).
Notice that those four exponentials are the dominant exponentials in the regions around the X-vertex.

An important feature of X-soliton solution is the existence of the \emph{constant} phase shift
which appears as a shift of the crest-line of each line-soliton at the interaction region.
For example, in the case of two solitons $[1,2]$ and $[3,4]$ in Figure~\ref{fig:PS}, 
each soliton has a negative phase shift $\Delta x_{[i,j]}$ which is determined 
only by the $k$-parameters,
equivalently, by the amplitudes and slopes of the solitons forming X-soliton.
In particular, for two solitons with equal amplitude $A_{[1,2]}=A_{[3,4]}=:A_0$
and slope $-\Psi_{[1,2]}=\Psi_{[3,4]}=:\Psi_0>0$, the phase shift is given by
\cite{CK:09}
\begin{equation}\label{phase-shift}
\Delta x_{[1,2]}=\Delta x_{[3,4]}=\frac{1}{\sqrt{2A_0}}\ln\left(1-\frac{2A_0}{\tan^2\Psi_0}\right)
\end{equation}
Figure \ref{fig:PS} illustrates the interaction with phase shifts $\Delta x_{[i,j]}$ 
with the chord diagram of $\pi=(2143)$ expressing two line-solitons $[1,2]$ and $[3,4]$.

\begin{figure}[h!]
\begin{center}
\raisebox{1in}{(a)}~\includegraphics[width=4.2cm]{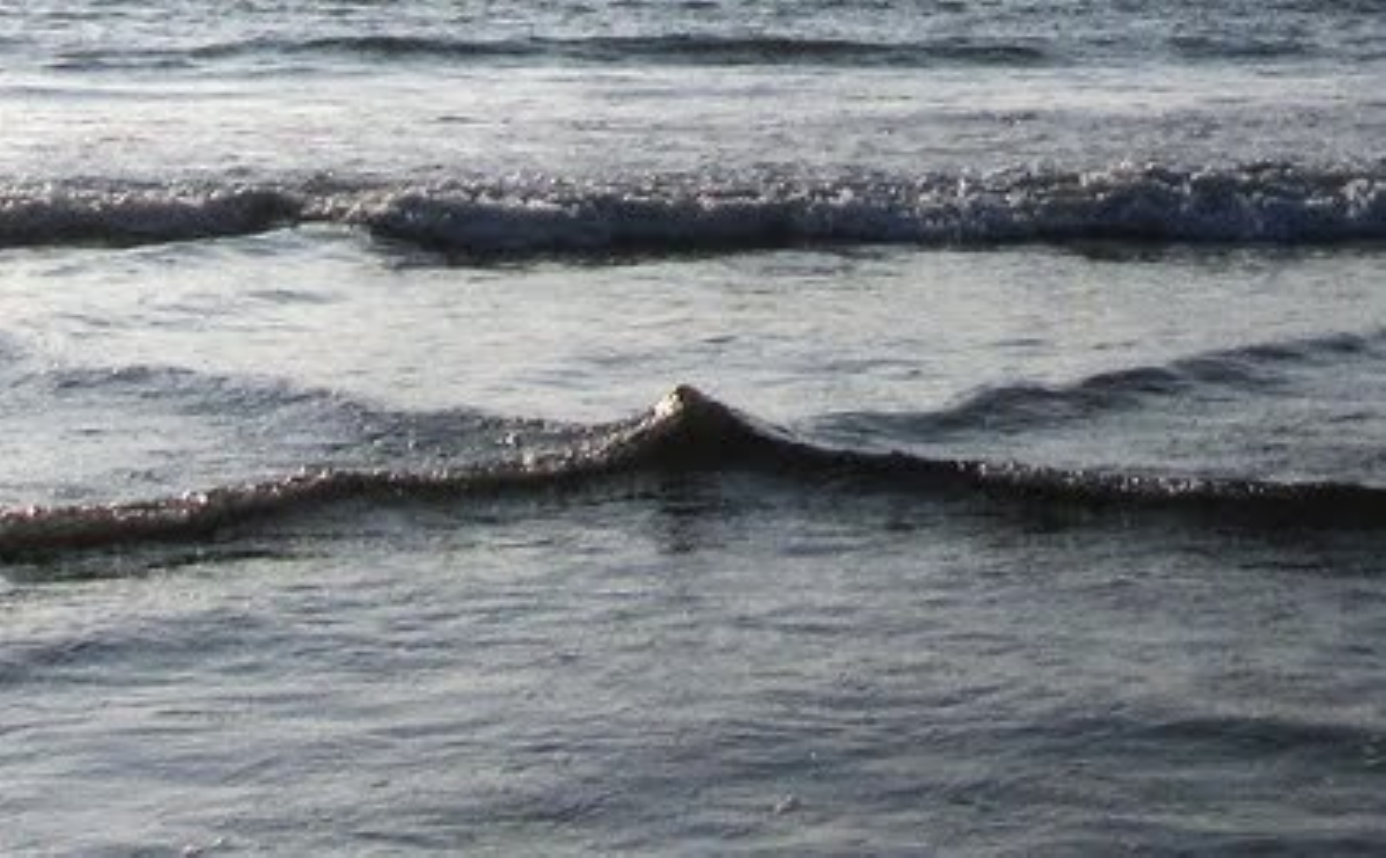} \hskip1.5cm
\raisebox{1in}{(b)}~\includegraphics[width=7.8cm]{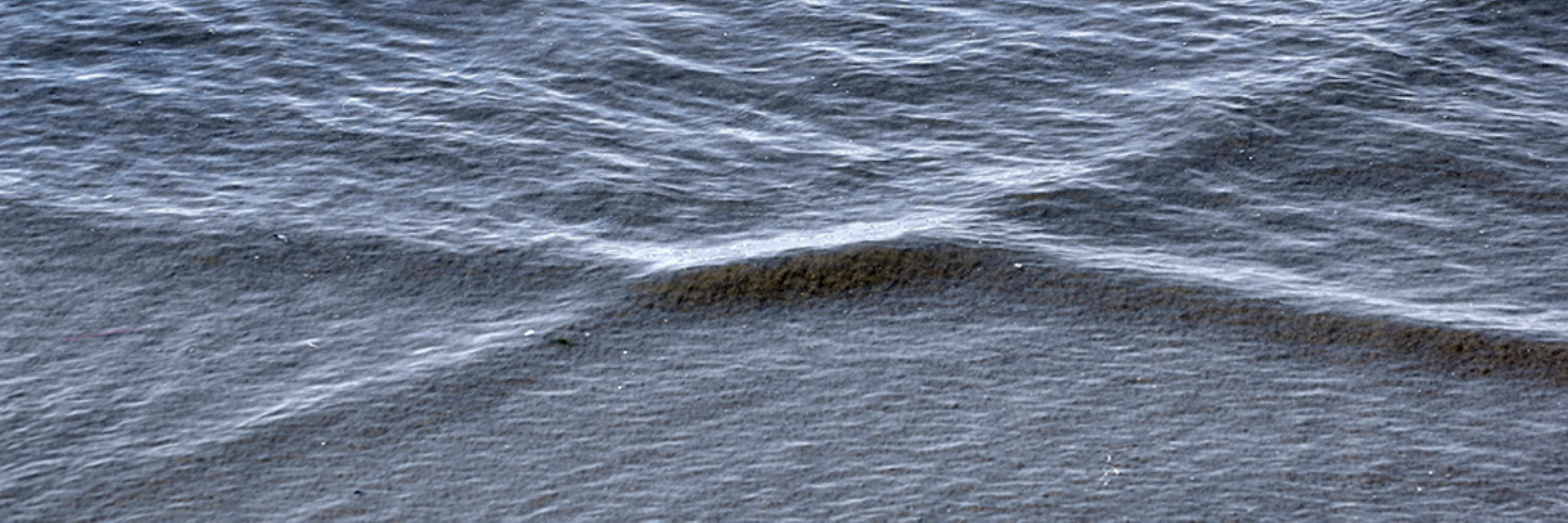} 
\end{center}
\caption{Photographs by M. J. Ablowitz (left panel) and 
D. E. Baldwin (right panel)~\cite{AB:12a}}
\label{fig:stem}
\end{figure}

It is quite important to note that in a real physical situation, the phase shift 
determined by the amplitudes and slopes of solitary waves forming X-shape pattern
should be of the same order of scaling as used in the derivation of the KP equation
from the Euler equations.
The scaling for the KP equation is given by the parameter
$\epsilon \sim a_0/h_0 \sim (h_0/\lambda_0)^2\sim \tan^2\psi_0$ where $a_0$ is the wave amplitude,
$h_0$ is the water depth, $\lambda_0$ is the wavelength
and $\psi_0$ is the slope of the soliton. The typical value of $\epsilon$ in real 
physical situation ranges from $10^{-2}$
to $10^{-1}$. If we take $\epsilon \approx 10^{-2}$, then
the phase shift can be at most twice the soliton wavelength.
In Figure~\ref{fig:PS}, we take $2A_0=\epsilon$ and $\tan^2\Psi_0=(1+2\cdot 10^{-4})\epsilon$ 
i.e., $\tan^2\Psi_0-A_0=\epsilon10^{-4}$ in \eqref{phase-shift}, in order to demonstrate
a large phase shift ($\sim$ 4 times the the soliton wavelength). Such large
phase shifts are highly atypical since a small difference of  
$\epsilon10^{-4} \approx \epsilon^3$ between
physical quantities which are of order $\epsilon$ can be easily destroyed by small
external perturbations or even by the higher order effects. 
(If $\epsilon\approx 10^{-1}$, then we need the accuracy of 
order $\epsilon^5= 10^{-4}\epsilon$).
In this sense, a long stem observed in a real wave pattern \emph{cannot} be
due to the phase shift. In other words, the wave pattern having a long stem
 does not correspond to an X-soliton of the KP equation.
For example, the wave pattern in Figure \ref{fig:stem}(a) shows
an X-type interaction, while the wave pattern in Figure \ref{fig:stem}(b)
is not of X-type, and the stem appearing in the middle of 
Figure \ref{fig:stem}(b) is an intermediate solitary wave
which interact resonantly with other two solitons 
at the trivalent vertices. In fact, we expect the wave pattern shown in
Figure \ref{fig:stem}(b) to be non-stationary (although records of its time evolution
is not available to us). In the framework of KP theory, the wave patterns in 
Figures~\ref{fig:stem} approximate two entirely different 
soliton solutions of the KP equation. 

\begin{Remark}
From \eqref{phase-shift}, one can see  that this type of X-soliton solution 
exists only if $2A_0<\tan^2\Psi_0$. 
The formula \eqref{phase-shift} breaks down at the critical slope 
$\tan^2\Psi_c:=2A_0$, whence a new soliton solution of KP emerges at this limit.
In this case, the X-vertex degenerates to a Y-vertex as the two solitons interact
resonantly and generate a third soliton, called the Mach stem~\cite{M:77b}.
One can also compute the maximum amplitude occurring at the mid-point of the interaction, 
and it is given by \cite{CK:09}
\[
u_{\text{max}}=\frac{4A_0}{1+\sqrt{\Delta_0}}\quad\text{with}\quad
\Delta_{0}:=1-\frac{2A_0}{\tan^2\Psi_0}.
\]
Then at the critical angle $\Psi_c$, the maximum amplitude can reach four times of the
solitons. This interaction phenomena is referred to as
the Mach reflection which has important application in the study of rogue waves. 
A recent study on the Mach reflection for KP solitons can be found in~\cite{LYK:11}.
In particular, it was found in~\cite{KOT:09} that the line-soliton solution of type
$\pi=(3142)$ can be used to describe the non-stationary wave patterns generated by the 
Mach reflection for the cases with $2A_0>\tan^2\Psi_0$, where the X-soliton becomes 
singular (see \cite{M:77b}).
\end{Remark}

\begin{figure}[h!]
\begin{center}
\includegraphics[width=14cm]{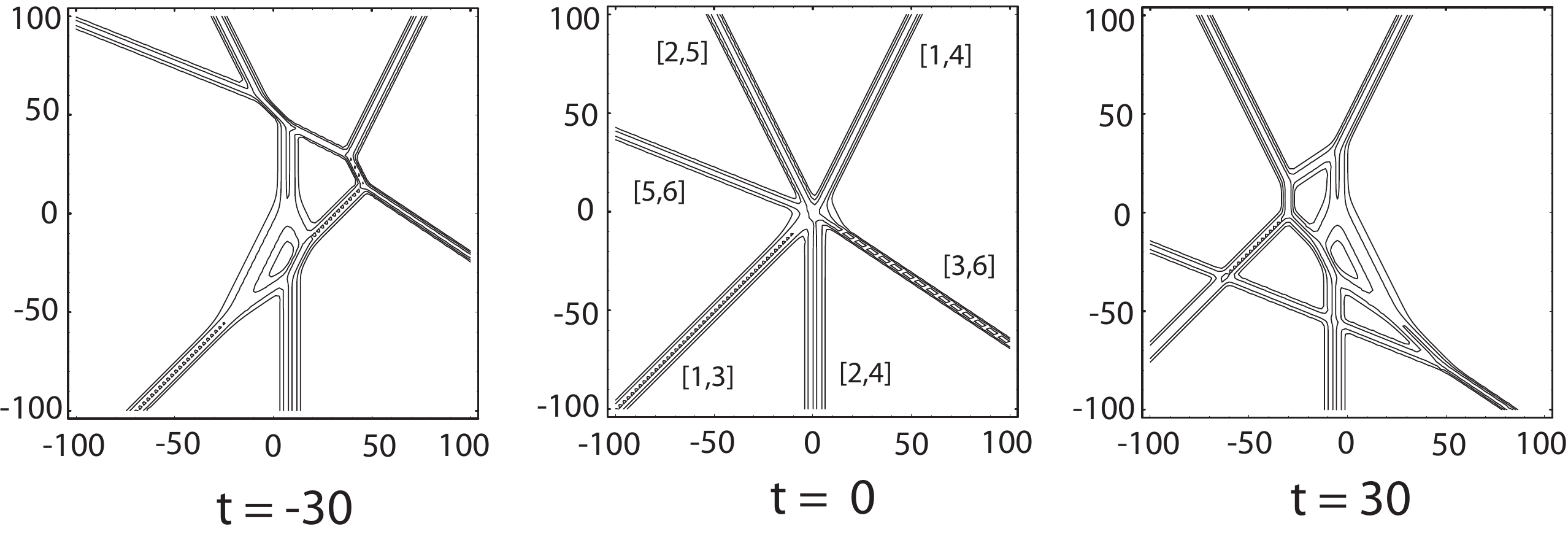}
\end{center}
\caption{Example of web-soliton solution with $\pi=\binom{1~2~3~4~5~6}{4~5~1~2~6~3}$.}
\label{fig:33soliton}
\end{figure}

A general line-soliton solution of the KP equation can be regarded as a collection
of one-solitons and Y-solitons which fit together to form a web-structure
with X- and Y-vertices in the interaction region
 as shown in Figure~\ref{fig:33soliton}.
An alternate combinatorial description of this solution can be obtained by
gluing the local chord diagrams corresponding to those component one-solitons and
Y-solitons and to form the complete chord diagram for the derangement $\pi$ 
associated with the general soliton solution. We will implement this technique of
``gluing'' of chords in the next section in order to reconstruct a KP soliton
solution from a {\it real} wave pattern and will illustrate this method
with explicit examples.

\section{Inverse problem}
In this section we discuss the problem of constructing an exact soliton solution
of the KP equation from a given wave pattern satisfying the assumptions required to 
derive the KP equation. In other words, the pattern should describe
interactions of small amplitude, quasi-two-dimensional waves of long wavelength traveling 
in one direction. Here we consider the surface wave patterns on shallow water
observed in long, flat stretches of ocean beaches.
Provided that the wave pattern approximates a KP soliton solution,
we assume that each component wave in the pattern approximates a $[i,j]$-soliton
for some index pair $i<j$.
Then the construction procedure consists of the following steps:
\begin{enumerate}
\item Set the $xy$-coordinates, so that all solitary waves in the pattern are 
traveling almost in the positive $x$-direction (towards the shore).
\item Using \eqref{ij} and \eqref{ijl}, find the $k$-parameters by measuring the 
amplitudes and slopes of the solitary waves in the pattern.
\item Define the derangement $\pi$ and its chord-diagram 
from the $k$-parameters obtained in the step 2.
\item Determine the form of the totally non-negative matrix $A$ 
from the derangement $\pi$ using combinatorial techniques~\cite{KW:12}.
\item  Using \eqref{theta}, find the matrix elements of $A$ from the minors
$\Delta_I(A)$ by measuring the locations of the solitary waves in the pattern.
\end{enumerate}

We remark that if there are resonances in the wave pattern, 
it is sufficient to measure the amplitudes, or slopes or  
locations of only a subset of the solitary wave forms in the pattern, rather
than the entire set.

The wave patterns that we consider for the inverse problem are assumed to be
{\it generic} in the sense that the topological structure of the pattern remains the
same over a finite period of time. That is, the number of X- and Y-vertices in the 
pattern remains the same over a certain time period.
If a generic wave pattern contains a line soliton
which interacts non-resonantly with all other waves (i.e. the interactions make
only X-vertices), then it can be  
treated as a separate $[i,j]$-soliton in the sense
that its chord-diagram (see Figure~\ref{fig:1soliton}) is disjoint (non-crossing) 
from the chord diagram of the remaining pattern. It is for this reason we only consider
wave patterns where each line soliton is incident on {\it at least} one trivalent vertex.
That is, each line soliton interacts resonantly with two other line solitons
forming a Y-vertex. 

\begin{figure}[h!]
\begin{center}
\includegraphics[width=13cm]{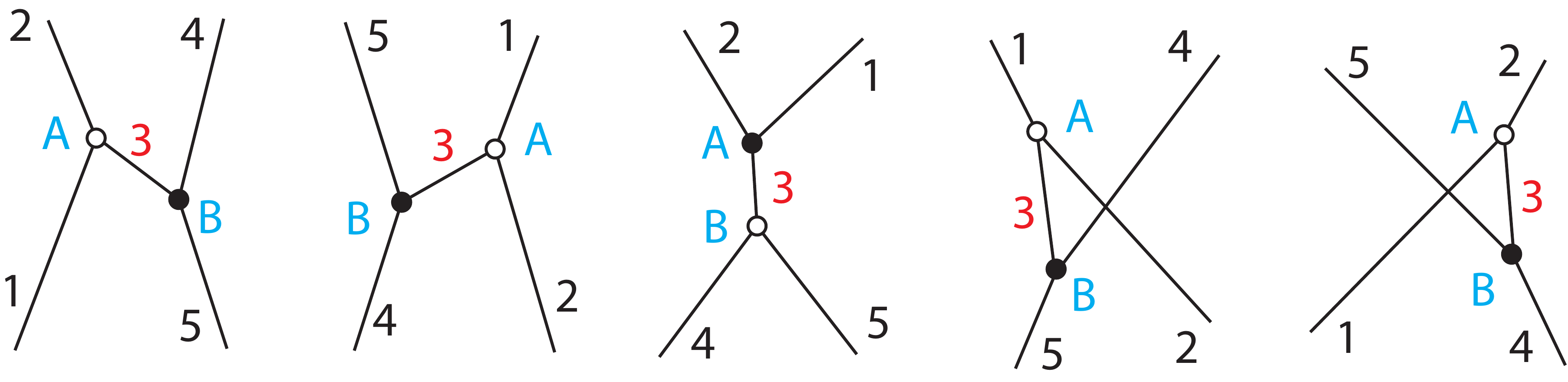}
\end{center}
\caption{Five distinct wave patterns obtained by gluing two Y-shape waves.}
\label{fig:2Res2}
\end{figure}

In this note,  we primarily consider wave patterns with only {\it two} trivalent
vertices consisting of two Y-type (one Y-shape and other \Y-shape)
waves which have a common intermediate solitary wave.
We distinguish the Y-type waves by marking their trivalent vertices, in particular,
we assign a white vertex for the \Y-shape wave, and a black vertex for the Y-shape wave.
Identifying the common intermediate wave from each Y-shape wave, one can see that
there are 5 distinct cases for such wave pattern 
as shown in Figure~\ref{fig:2Res2}.  
The resonant vertices in each of these patterns are labeled $\mathsf{A}$ 
(upper vertex) and $\mathsf{B}$ (lower vertex). The solitary waves incident at the 
vertex $\mathsf{A}$ are numbered as 1, 2 and 3, while the solitary waves
numbered as 3, 4 and 5 are incident at the vertex $\mathsf{B}$. Thus each pattern consists
of two Y-type waves glued together by the common solitary wave 3 joining vertices
$\mathsf{A}$ and $\mathsf{B}$. Besides serving as simple examples to illustrate our 
method, these patterns are
observed fairly regularly in flat ocean beaches. However, such wave patterns
{\it cannot} be modeled by Hirota's 2-soliton formula which 
gives a stationary pattern, and contains only
one free parameter determining the phase shift due to  
non-resonant interaction of the $X$-soliton described in Section~\ref{sec:X}.
The exact line-soliton solutions corresponding to these resonant wave patterns
are \emph{non}-stationary and contain more than one free parameters
(a complete classification of the line-soliton solutions corresponding to those patterns in
Figure \ref{fig:2Res2} is given in \cite{CK:09}).

In order to construct KP solitons from those patterns, we first
identify each solitary wave with an $[i,j]$-soliton for some parameters $k_i<k_j$.
Let us assign $\{k_a,k_b,k_c\}$ at the vertex $\mathsf{A}$, and  $\{k_{a'},k_{b'},k_{c'}\}$
at the vertex $\mathsf{B}$. Thus each one-soliton can be parametrized as follows:
\begin{center}
\begin{tabular}{ccccc}
Soliton 1 & Soliton 2 & Soliton 3 & Soliton 4 & Soliton 5 \\
$\{k_a,k_b\}$ & $\{k_a,k_c\}$ & $\{k_b,k_c\}, 
\{k_{b'},k_{c'}\}$ & $\{k_{a'},k_{b'}\}$ & $\{k_{a'},k_{c'}\}$ 
\end{tabular}
\end{center}
Note that there are two parametrizations (with respect to the two vertices A and B)
for the intermediate soliton 3 which is common to the Y-solitons.
Let $\tan \Psi_i, \, i=1,2,\ldots,5$ denote the slope of each one-soliton.
In particular, the slope of soliton 3 is given as $\tan\Psi_3 = k_b+k_c = k_{b'}+k_{c'}$.
Then the $k$-values can be solved uniquely from the slope measurements
at the vertices $\mathsf{A}$ and $\mathsf{B}$ for each Y-soliton as given by \eqref{ijl}, i.e.
\begin{subequations}\begin{align}
&\left\{\begin{array}{lll}
k_a=\half(\tan\Psi_1+\tan\Psi_2-\tan\Psi_3)\,,  \\[0.5ex]
k_b=\tan\Psi_1-k_a\,, \quad k_c=\tan\Psi_2-k_a,
\end{array}\right. \label{2Ya}\\[1.0ex]
&\left\{\begin{array}{lll}
 k_{a'}=\half(\tan\Psi_4+\tan\Psi_5-\tan\Psi_3)\,, \\[0.5ex]
k_{b'}=\tan\Psi_4-k_{a'}\,, \quad k_{c'}=\tan\Psi_5-k_{a'}
\end{array}\right.
\label{2Yb}
\end{align}
\end{subequations} 
Without loss of generality, the two sets of $k$-parameters for soliton 3 can be
ordered as $k_b < k_c$ and $k_{b'} < k_{c'}$. If the given wave pattern is an exact
KP soliton, then one should have $k_b=k_{b'}$ and $k_c=k_{c'}$.
But from equations \eqref{2Ya} and \eqref{2Yb} we have
\[k_b-k_{b'} = - (k_{c}-k_{c'})= \half[(\tan\Psi_1-\tan\Psi_2)-(\tan\Psi_4 - \tan\Psi_5)].
 \]
This has a non-zero value which measures the deviation from the KP soliton
and is also due to the error in the measurement. 
To approximate a real wave pattern by a KP soliton, 
it is reasonable to prescribe the $k$-parameters
of soliton 3 by taking the average of the two sets of $k$-values.
Therefore, we define the $k$-parameters of soliton 3 as
\[
\tilde{k}_b=\half(k_b+k_{b'}), \qquad \quad \tilde{k}_c=\half(k_c+k_{c'}) \,,
\] 
which preserves the slope of soliton 3, i.e., $\tan \Psi_3 = \tilde{k}_b+\tilde{k}_c$,
and leads to the following re-parametrization of the line solitons:
\begin{center}
\begin{tabular}{ccccc}
Soliton 1 & Soliton 2 & Soliton 3 & Soliton 4 & Soliton 5 \\
$\{k_a,\tilde{k}_b\}$ & $\{k_a,\tilde{k}_c\}$ & $\{\tilde{k}_b,\tilde{k}_c\}$ & 
$\{k_{a'},\tilde{k}_b\}$ & $\{k_{a'},\tilde{k}_c\}$
\end{tabular}
\end{center}

We remark here that the conditions $k_b=k_{b'}$ and $k_c=k_{c'}$, are consequences 
of the fact that the amplitudes $A_{[b,c]}=A_{[b',c']}$ in the exact KP theory.  
In fact, from the amplitude formula in \eqref{ij}, we have
\[k_b-k_{b'} = - (k_{c}-k_{c'})= 
\frac{1}{2}\left(\sqrt{2A_{[b,c]}}-\sqrt{2A_{[b',c']}}\right) \,,  \]
which vanishes when $A_{[b,c]}=A_{[b',c']}$. 
One should however note that the amplitude along a solitary wave in the pattern
may not be uniform giving rise to different measurements at the vertices A and B.
Moreover, the amplitude formula $A_{[i,j]}=\half(k_i-k_j)^2$ is only correct
up to first order, and one needs higher order corrections to obtain the corresponding 
KP amplitude~\cite{LYK:11} (see also Lecture 1 in \cite{K:13}).
On the other hand, the slope of each solitary wave in the observation is the same as that 
used in the KP equation. For this reason, we determine the $k$-parameters from the slope
measurements only.

In what follows, we provide a step by step algorithm to reconstruct the exact KP soliton
solution from a given wave pattern with two trivalent resonant vertices.

\subsection{Algorithm for the inverse problem}\label{sec:algorithm}
\begin{figure}[h!]
\begin{center}
\includegraphics[width=13cm]{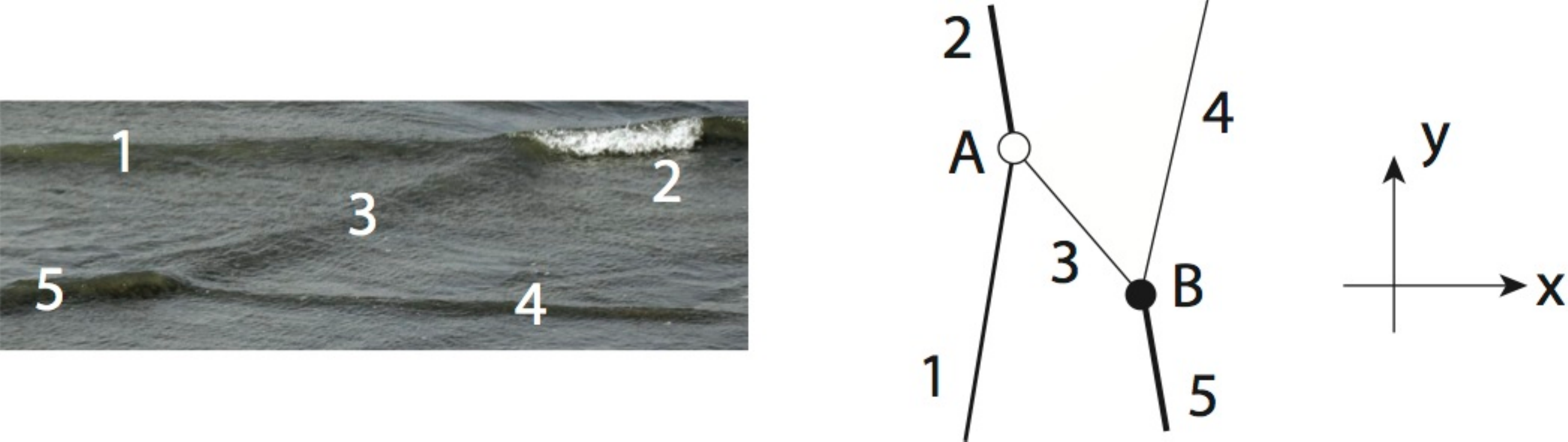}
\end{center}
\caption{A real example of the wave pattern corresponding to the
first pattern in Figure \ref{fig:2Res2}.  Photograph of the wave pattern 
taken by D. E. Baldwin~\cite{AB:12a}}
\label{fig:Mark1}
\end{figure}

For specificity, we use a real example of the wave pattern 
in Figure \ref{fig:Mark1} which corresponds to the first pattern
in Figure~\ref{fig:2Res2}. 

\subsubsection*{Step 1: Finding the $k$-parameters}
We first set the $xy$-coordinates, so that the wave is propagating along almost 
$x$-direction. We choose the origin as the vertex $\mathsf{A}$ in Figure~\ref{fig:Mark1}. 
Then we trace the wave crests as a graph, and measure the angles of all the 
line solitons with respect to the positive $y$-axis.
For this example, we estimate the angles for the line solitons to be  
\[
\Psi_1=-15^{\circ},\quad \Psi_2=3^{\circ}, \quad \Psi_3=25^{\circ}, \quad
\Psi_4=-13^{\circ},\quad\Psi_5=~5^{\circ}.
\]
Next from \eqref{2Ya} and \eqref{2Yb}, one obtains
$k_a=-0.341,k_b=0.073$ and $k_c=0.393$ at the vertex $\mathsf{A}$, and
$k_{a'}=-0.305,k_{b'}=0.074$ and $k_{c'}=0.392$ at the vertex $\mathsf{B}$.
After averaging to get $\tilde{k}_b,\tilde{k}_c$ and then arranging the
$k$-parameters in increasing order gives
\[
(k_a,k_{a'},\tilde{k}_b,\tilde{k}_c)= (k_1,k_2,k_3,k_4)=(-0.341,-0.305,0.0735,0.3925)\,.
\]
Note that the slope data at the resonant vertices is sufficient to obtain the
$k$-parameters, the amplitude data for the solitons, which is difficult to
measure from a photograph, is not necessary in this case.
Note that the amplitude of each one-soliton can be calculated from the $k$-values
using \eqref{ij}. Here we see that the soliton 2 has the maximum 
amplitude, $A_{[1,4]}=\half(k_4-k_1)^2=0.269$, and the soliton 3 has the maximum 
slope, $\tan\Psi_{[3,4]}=k_3+k_4=0.466$.  These are rather large values for the 
KP approximation, and we may need higher order corrections
to obtain a better result (see Lecture 1 in \cite{K:13}).

\begin{figure}[h!]
\begin{center}
\includegraphics[width=11.5cm]{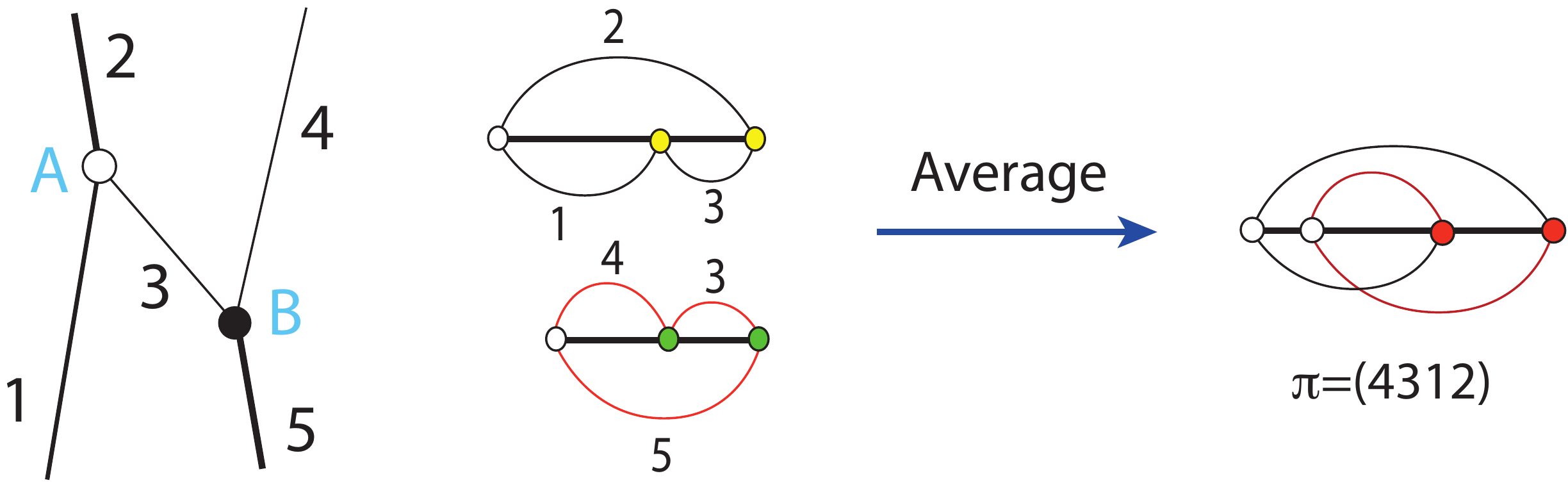}
\end{center}
\caption{Chord gluing algorithm for the wave pattern in Figure \ref{fig:Mark1}}
\label{fig:Mark2}
\end{figure}

From this set of $k$-parameters, the procedure to find the derangement $\pi$ for the data may be 
illustrated as a process of gluing the chord diagrams of the two Y-solitons with
vertices $\mathsf{A}$ and $\mathsf{B}$ through the common chord as illustrated in Figure~\ref{fig:Mark2}.  
At the vertex $\mathsf{A}$, one has a \Y-soliton of the type 
formed by line solitons labeled 1,2 and 3, and the corresponding chord diagram 
is indicated by the top diagram in the
middle of Figure~\ref{fig:Mark2}. The dots in this diagram denote the $k$-parameters
$k_a<k_b<k_c$.  Similarly, the bottom chord diagram 
corresponds to the Y-soliton
at the vertex $\mathsf{B}$ formed by solitons 3,4 and 5 with the dots marking
the $k$-parameters $k_{a'}<k_{b'}<k_{c'}$. Notice that the common chord (labeled 3)
in these two diagrams correspond to soliton 3 joining the vertices $\mathsf{A}$ and $\mathsf{B}$. 
We then superpose the top and bottom diagrams and ignore the common chord for 
the intermediate soliton 3 and recover the complete chord diagram of the asymptotic
solitons shown by the rightmost diagram in Figure~\ref{fig:Mark2}. In this diagram
the $k$-parameters are marked (by the 4 dots) as $k_a=k_1, k_{a'}=k_2$ and the
average values $\tilde{k}_b=k_3, \tilde{k}_c=k_4$ with $k_1<k_2<k_3<k_4$. 
This diagram corresponds to the derangement 
$\pi=\left(\begin{smallmatrix}1&2&3&4\\4&3&1&2\end{smallmatrix}\right)$,
or simply as $\pi=(4312)$ which is the one-line notation of permutation.
The top chords in this diagram are associated to the asymptotic solitons 2 and 4
for $y \gg 0$ which are identified as $[1,4]$-
and $[2,3]$-soliton. Similarly, the bottom 
chords in this diagram
are associated to the $[1,3]$- and $[2,4]$- asymptotic line solitons for $y \ll 0$.
This is summarized in Figure \ref{fig:Mark2}.

\subsubsection*{Step 2: Finding the form of the $A$-matrix}
Recall that in order to construct the $\tau$-function in \eqref{tau}
for the soliton solutions, one needs the $N \times M$ totally nonnegative matrix $A$
in addition to the $k$-parameters. The $A$-matrix can derived from the derangement
$\pi$ obtained in Step 1 by employing combinatorial methods based on Deodhar
decomposition of Grassmann varieties. This has been developed in a recent paper
by Kodama and Williams~\cite{KW:12}, we refer the interested readers to that article
and omit the details in this short note. Alternatively, the interested reader may also
consult the work of Chakravarty and Kodama~\cite{CK:08, CK:09} where in particular, 
the $A$-matrices for those patterns in Figure \ref{fig:2Res2} were explicitly given 
in reduced row echelon form for 
the corresponding derangements $\pi$. For the current example with $\pi=(4312)$, the
$A$-matrix is given by either
\[
A=\begin{pmatrix} p_3p_4 & p_4 &  0  & - 1\\ 0  &  p_1  &  1  &  0 \end{pmatrix}
\quad \text{or}\quad \begin{pmatrix} 1 & 0 & -b & -c\\0 & 1 & a & 0\end{pmatrix},
\]
with three real parameters $\{p_1,p_3,p_4\}$ or $\{a,b,c\}$.
The second matrix (given in \cite{CK:09}) is the reduced row echelon form of 
the first one.
Note that $A$ is a totally nonnegative matrix, i.e., $\Delta_I(A) \geq 0$ for all
$I \in \binom{[M]}{N}$, 
if all $p_1,p_3,p_4>0$ (or all $a,b,c>0$). In this case, the corresponding exact soliton 
solution of KP is non-singular for all $x, y$ and $t$.  We also remark that the $\tau$-function
generated by the matrix $A$ contains five exponential terms, i.e.
\[
\tau =\sum_{1\le i<j\le 4}\Delta_{i,j}(A)E_{i,j}\qquad \text{with}\quad \Delta_{i,j}(A)\ne0~~\text{except}~~\Delta_{1,4}(A)=0.
\]
One should compare this with the case of X-soliton where the $\tau$-function has
four exponential terms.  Notice that if $p_4=0$ (or $b=0$) with keeping $p_3p_4=a$ and $p_1=c$,
then the corresponding X-soliton consists of $[1,4]$- and $[2,3]$-solitons with the phase shift,
i.e. the X-soliton is of the type $\pi=(4321)$.

\subsubsection*{Step 3: Determining the values of the $A$-matrix}
The last step is to explicitly find the entries of the $A$-matrix.
This is accomplished by evaluating the $p$-parameters from the given wave pattern.  

\begin{figure}[h!]
\begin{center}
\includegraphics[width=13cm]{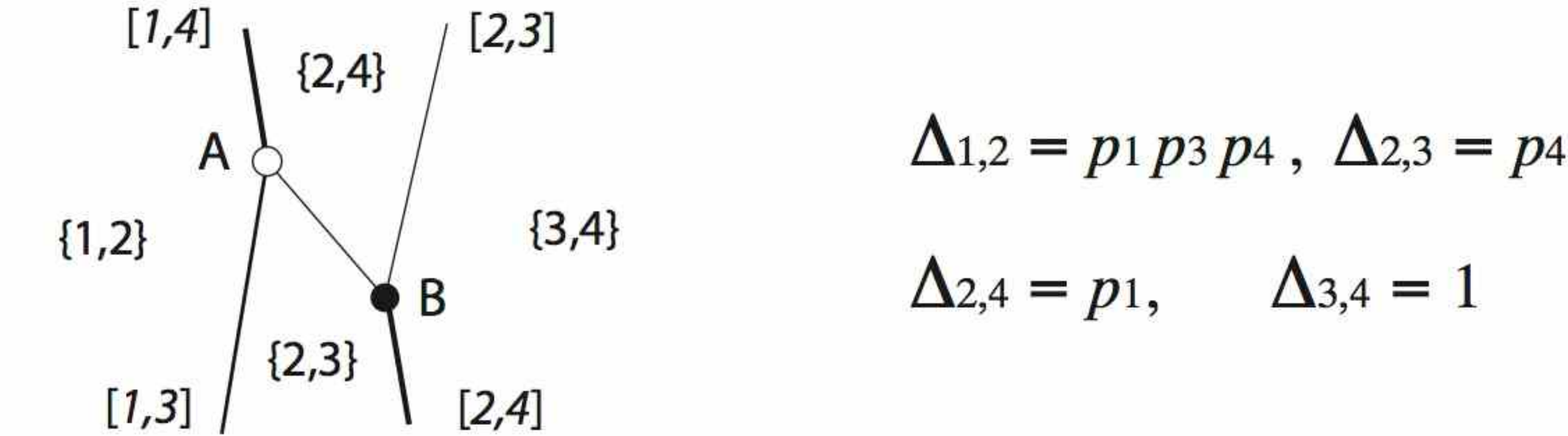}
\end{center}
\caption{Dominant exponentials and the minors of the matrix $A$ for
the wave pattern in Figure \ref{fig:Mark1}}
\label{fig:Plucker}
\end{figure}

Figure~\ref{fig:Plucker} shows the wave pattern of our example with all the
$[i,j]$ asymptotic solitons labeled. The index set $I=\{i,j\},\, i<j$ indicates 
each region in the $xy$-plane where the exponential $E_I$ is dominant. 
The corresponding minor $\Delta_I(A)$ of the matrix $A$ with column set $I$ 
is given in 
terms of the $p$-parameters. Recall from the previous section that the equation 
of the line corresponding to the $[i,j]$-soliton is given by $\Theta_{[i,j]}=0$  
for a fixed value of $t$, where $\Theta_{[i,j]}$ is given by \eqref{theta}.
Notice that the expression for $\Theta_{[i,j]}$ contains the constant $x_{[i,j]}$
which involves the ratios of the minors $\Delta_I(A)$ and $\Delta_J(A)$
where $I$ and $J$ label the adjacent regions in the $xy$-plane 
whose common boundary is the line corresponding to the $[i,j]$-soliton. 

We pick a point $(x_0,y_0)$ on the line
corresponding to the $[2,4]$-soliton in Figure~\ref{fig:Plucker}. Setting the
time $t=0$, we have the equation of this line,
\[
x+(k_2+k_4)y+\frac{1}{k_2-k_4}\ln\frac{\Delta_{2,3}(A)(k_3-k_2)}{\Delta_{3,4}(A)(k_4-k_3)}=0 \,. 
\]
Using the facts that $\Delta_{3,4}(A)=1$ and the point $(x_0,y_0)$ is on the line, one 
obtains
\[
\Delta_{2,3}(A)=p_4=\frac{k_4-k_3}{k_3-k_2}\exp\left((k_4-k_2)x_0+(k_4^2-k_2^2)y_0\right) \,.
\] 
Similarly, selecting a point $(x_0,y_0)$ on line corresponding to the $[2,3]$-soliton, yields 
\[
\Delta_{2,4}(A)=p_1=\frac{k_4-k_3}{k_4-k_2}\exp\left((k_3-k_2)x_0+(k_3^2-k_2^2)y_0\right) \,,
\]
and finally, by choosing a point on either the $[1,4]$- or the $[1,3]$-line, one can get 
$\Delta_{1,2}(A)$, which in turn, yields $p_3$. 
In this way, we can evaluate the $p$-parameters (hence the $A$-matrix) from the graph 
corresponding to a photograph of the wave pattern after we have identified all 
the $[i,j]$-solitons in the pattern. 

\begin{figure}[h!]
\begin{center}
\includegraphics[width=14cm]{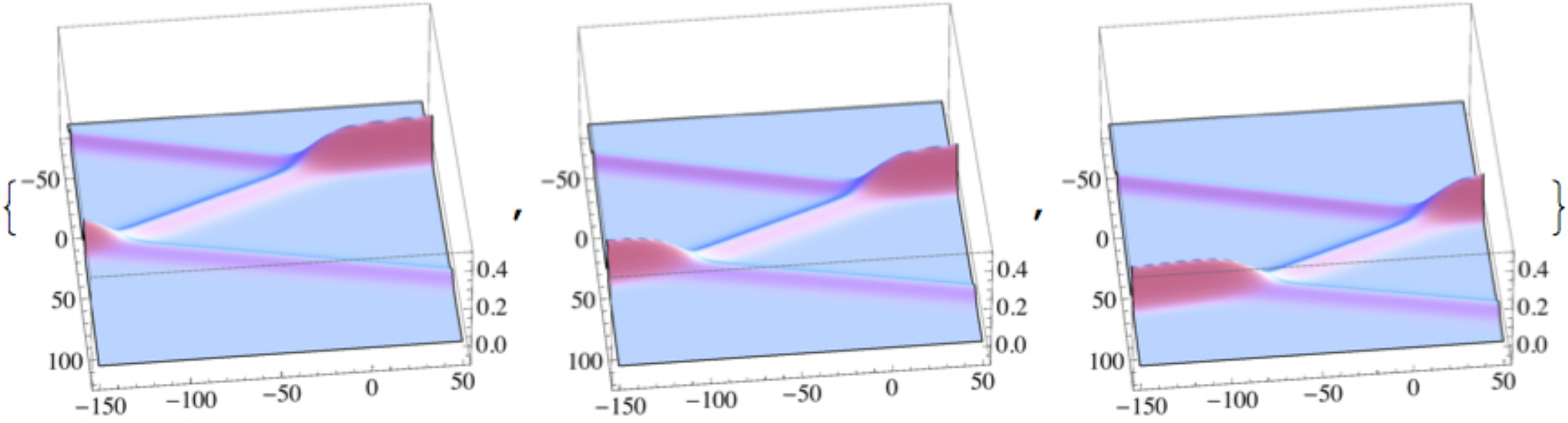} 
\end{center}
\caption{Evolution of the KP soliton constructed from the wave pattern
in Figure \ref{fig:Mark1}.}
\label{fig:Mark-Ex2}
\end{figure}

For this example, we chose the origin as the vertex $\mathsf{A}$ common to lines
corresponding to the $[1,3]$- and $[1,4]$-solitons, and 
estimate the location of the vertex $\mathsf{B}$ as $(50, -107.2)$
which lie on the $[2,3]$- and $[2,4]$-soliton lines. Taking $(x_0,y_0)$ as those points,  
we obtain the values
$p_1=9.1 \times 10^{11},  \,p_3=1.1 \times 10^{-11},\, p_4=16.9 \times 10^{11}$ 
which gives the exact $A$-matrix in Step 2. From this $A$-matrix and the $k$-parameters 
already found in Step 1, one can now construct the $\tau$-function and the KP soliton
solution using \eqref{u}. Plots of the exact solution, $u=2(\ln\tau)_{xx}$, which approximates
the snapshot of the original wave pattern (Figure~\ref{fig:Mark1}),
are shown in Figure~\ref{fig:Mark-Ex2}. The middle plot at $t=0$
corresponds to the photograph of the wave pattern in Figure~\ref{fig:Mark1}, 
the left and right plots are for $t<0$ and $t>0$, respectively. 
Notice that the wave pattern is not stationary, and the stem (soliton 3) is getting shorter 
which means it is a an intermediate 
soliton rather than a stationary phase shift. 

\section{Examples}
Here we present a few more examples to illustrate our method.
In these examples we consider a succession of photographs of an evolving
wave pattern and compare the dynamics with the time evolution of the exact solution
reconstructed by the algorithm described in Section \ref{sec:algorithm}.

\subsection{Example 1: An Estonian beach}
\begin{figure}[h!]
\begin{center}
\includegraphics[width=12cm]{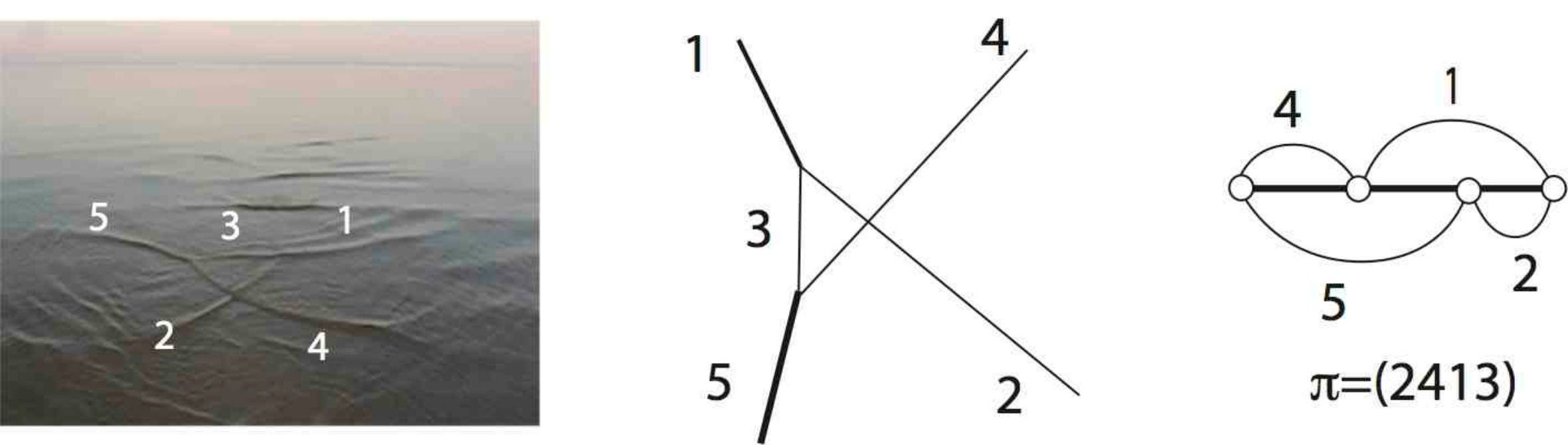}
\includegraphics[width=15cm]{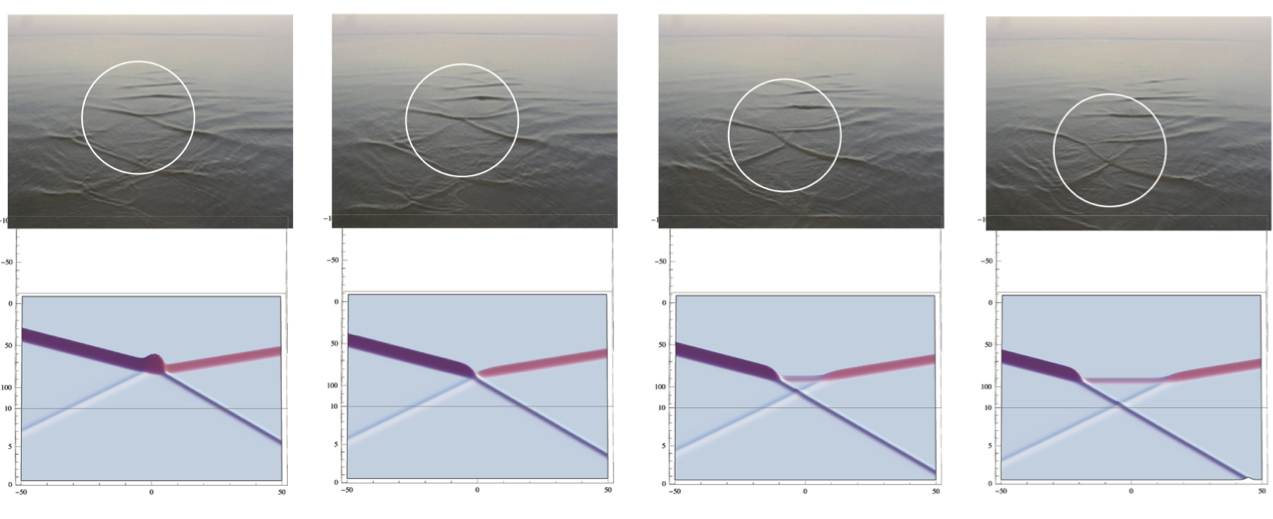}
\end{center}
\vskip-0.2in
\caption{Waves in Lake Peipsi, Estonia. 
 Snapshots from a video courtesy 
Ira Didenkulova.\label{fig:Est-ex}}
\end{figure}

The top panel in Figure~\ref{fig:Est-ex} shows a snapshot of a wave pattern
observed on a beach of Lake Peipsi, Estonia.  
This photograph is part of a video footage
capturing the dynamics of the wave pattern over a period of time.
This corresponds to the wave pattern of the fourth case 
in Figure \ref{fig:2Res2} with two Y-vertices and an X-vertex.  We then apply our
inverse problem algorithm to this snapshot, which is chosen to be $t=0$, to construct 
the exact solution $u(x,y,0)$, then compare $u(x,y,t)$ for $t \neq 0$ with other
snapshots taken from the video footage. 
The middle picture of the top panel in Figure~\ref{fig:Est-ex} shows the graph traced from
the wave crests in the pattern. The solitary waveforms in the pattern
are not straight lines, we first locate the vertices and join them to obtain the lines
marked 2,3 and 4. The lines 1 and 5 are traced by using the slopes computed
at each of their respective vertices. 
From the trace the angles of the line solitons are estimated to be 
$\Psi_1=14^\circ,\, \Psi_2=37^{\circ}, \, \Psi_3=0^\circ, \,\Psi_4=-42^\circ,
\Psi_5=-20^{\circ}$.  Following Step 1 in Section \ref{sec:algorithm}, we obtain 
the  $k$-parameters as $(k_1,k_2,k_3,k_4)=(-0.632,-0.268,0.268,0.517)$.  
Then we have the derangement $\pi=(2413)$ as shown in the top right panel 
of Figure \ref{fig:Est-ex}. Here solitons 2 and 5 correspond to the $[3,4]$- and $[1,3]$-solitons 
respectively, for $y \ll 0$, whereas solitons 1 and 4 correspond to
$[2,4]$- and $[1,2]$-solitons for $y \gg 0$. Finally, we construct the exact solution
following the prescription in Steps 2 and 3. The bottom two panels of Figure~\ref{fig:Est-ex} 
exhibit the comparison between the time evolutions of the exact solution and the actual 
wave pattern. The last snapshot in the middle panel is the snapshot
corresponding to $t=0$ from which the exact KP solution was constructed, the 
previous snapshots in the middle panel correspond to $t < 0$. 
Also note that the leftmost snapshot corresponds to the wave pattern
of the third case in Figure \ref{fig:2Res2} indicating that the evolving solution
can coincide with more than one configuration shown in Figure \ref{fig:2Res2}.

\subsection{Example 2: An Mexican beach}
\begin{figure}[h!]
\begin{center}
\includegraphics[width=16cm]{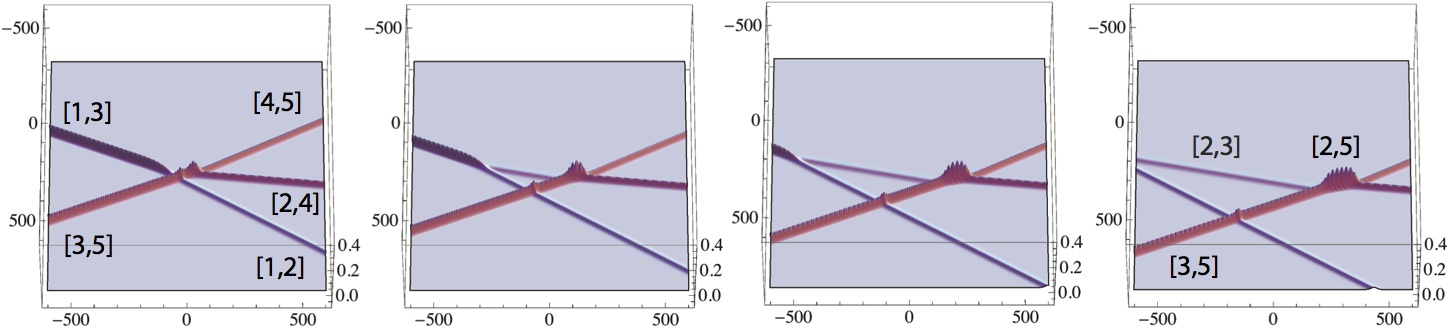}
\end{center}
\caption{Evolution of the KP soliton constructed from the wave pattern in 
Figure \ref{fig:f120015}}
\label{fig:MarkOr}
\end{figure}

Our next example is from
Figure \ref{fig:f120015} in Section 1 showing a wave pattern with three
line solitons for $y \gg 0$ and two line solitons for $y \ll 0$. 
The left photograph in the figure
was used to reconstruct the exact solution at $t=0$, the right photograph is 
of the same wave pattern taken at some $t>0$. Without giving the details we 
simply point out that in this case our algorithm  identifies the solitons
(from right to left) $[1,2]$, $[2,4]$ and $[4,5]$ for $y \gg 0$, and solitons
$[1,3]$ and $[3,5]$ for $y \ll 0$, from which we have the derangement $\pi=(24153)$.  
Let us describe the evolution of the
exact solution in Figure \ref{fig:MarkOr} in more details. 
There are several things to notice here. 
This solution has three Y-vertices and one X-vertex (these are more clear 
in the second and third frames). 
The solitons $[2,4]$ and $[4,5]$ interact at the rightmost ($y>0$) Y-vertex to 
form a large amplitude stem $[2,5]$-soliton which is visible in both
photographs in Figure \ref{fig:f120015} but more pronounced in the right
photograph. This clearly suggests that the $[2,5]$-soliton is not a phase shift
since it is a non-stationary stem arising from a resonant interaction of the solitons 
$[2,4]$ and $[4,5]$ from the right ($y\gg0$).

This $[2,5]$-soliton interacts resonantly with the 
$[3,5]$-soliton coming from the left ($y \ll 0$) at the second Y-vertex to form a
small amplitude $[2,3]$-soliton which is barely discernible in the first
frame as well as the wave pattern photograph in Figure \ref{fig:f120015}.
But as the pattern evolves the $[2,3]$-soliton is clearly visible in the second, third 
and fourth frames in Figure \ref{fig:MarkOr} as well as the right photograph of the 
wave pattern in Figure \ref{fig:f120015} taken at $t>0$. Again, this clearly
shows that the length of the $[2,3]$-soliton grows with time.
The $[2,3]$-soliton intersects with the $[1,2]$-soliton
on the right ($y \gg 0$) and the large amplitude asymptotic $[1,3]$-soliton on 
the left ($y \ll 0$) at the third Y-junction (seen on the left, in the second and 
third frames) to form a resonant triplet.

\section{Conclusion}
In this note we described an explicit algorithm to construct an exact 
solution of the KP equation approximating a given pattern of small amplitude,
long wavelength and primarily unidirectional shallow water waves.
This can be regarded as an ``inverse problem" in the sense that by
measuring some metric data such as the angles and locations of the solitary waves in the given
pattern with respect to a fixed
reference frame, it is possible to determine the data necessary to construct
the $\tau$-function associated with a KP line-soliton solution.
We have illustrated our method by applying it
to photographs and videos of real wave patterns. The constructed
exact solutions compare fairly well with the snapshots 
(at a fixed $t$) but more importantly we have shown that their time 
evolution is also in good agreement with the dynamics of these non-stationary patterns.
These exact solutions of the KP equation found in Sections 3 and 4 were derived
and classified in previous works by the authors, and they cannot 
be obtained from a simple Hirota $N$-soliton formula.

An important feature of this algorithm is that one can directly read off the
parameters for the exact soliton solution from the wave pattern data without
resorting to any ad hoc techniques to ``guess'' the soliton parameters
from the wave pattern.
One could also incorporate the measurement of amplitudes of
the wave forms in the algorithm although the measurement of the slopes are generally 
more accurate than that of the amplitudes of the wave forms in the pattern. The
possible inaccuracies in amplitude measurement are due to, for example,
the breaking of the waves and the higher order corrections to
the leading order KP equation. Moreover, it is difficult (if not impossible)
to measure the amplitudes if one were to use a photograph (or a video) of the wave pattern.
The slopes of the line solitons should be measured at a vertex (intersection
points of the interacting solitons), since the solitary wave form may not
be exactly a straight-line e.g., due to non-constant depth of water on a beach.
Our algorithm can be extended in principle to a wave pattern with arbitrary number
of solitary waves for $y \gg 0$ and $y \ll 0$ although such patterns may be
unstable and break up shortly after it is formed due to external perturbations.
  
\section*{Acknowledgements}
The authors are grateful to Mark Ablowitz, Douglas Baldwin and Ira Didenkulova
for the photographs and the video used in this article.
This research was supported by NSF grants DMS-1108694 (SC) and DMS-1108813 (YK).


\end{document}